\documentclass{ws-ijmpa}

\begin{document}

\newcommand{\rt}{\rightarrow}
\newcommand{\etal}{\it et al.\rm}

\newcommand{\jpsi}{J/\psi}
\newcommand{\psip}{\psi(2S)}
\newcommand{\ppp}{\pi^+\pi^-\pi^0}
\newcommand{\ar}{\rightarrow}
\newcommand{\uu}{\mu^+\mu^-}
\newcommand{\ppj}{\pi^+\pi^-J/\psi}
\newcommand{\rar}{\rightarrow}

\newcommand{\J}{J/\psi}
\newcommand{\ppbar}{p \bar{p}}
\newcommand{\pp}{\pi^+\pi^-}
\newcommand{\ra}{\rightarrow}
\newcommand{\Jpi}{\pi^0 J/\psi}
\newcommand{\Jeta}{\eta J/\psi}
\newcommand{\Jpipi}{\pi^0\pi^0 J/\psi}
\newcommand{\gx}{\gamma\chi_{c1,c2}}
\newcommand{\ggee}{\gamma\gamma e^+e^-}
\newcommand{\ggmm}{\gamma\gamma \mu^+\mu^-}
\newcommand{\ggll}{\gamma\gamma l^+l^-}
\newcommand{\MgJ}{M_{\gamma_h,J/\psi}}
\newcommand{\Mgg}{M_{\gamma\gamma}}
\newcommand{\ptochic}{\psi(2S)\ra \gamma\chi_{c1,2}}
\newcommand{\gguu}{\gamma\gamma\mu^+\mu^-}
\newcommand{\pppp}{\psi(2S) \rt \pi^+ \pi^- J/\psi}

\markboth{Frederick A. Harris}
{$J/\psi$ Decays and Charmonium Transitions}

%
\catchline{}{}{}{}{}
%

\title{BES RESULTS ON $J/\psi$ DECAYS AND \\
CHARMONIUM TRANSITIONS
}

\author{\footnotesize FREDERICK A. HARRIS \\
(For the BES Collaboration)}

\address{Department of Physics and Astronomy, University of Hawaii,
  2505 Correa Rd.\\
Honolulu, Hawaii  96822,
USA
}

\maketitle


\pub{Received (20 July 2004)}{ }

\begin{abstract}

Results are reported based on samples of 58 million $\jpsi$ and 14 million
$\psip$ decays obtained by the BESII experiment.  Improved branching
fraction measurements are determined, including branching fractions for
$\jpsi\to\ppp$, $\psip\ra \pi^0\J$, $\eta\J$, $\pi^0 \pi^0 J/\psi$,
anything $J/\psi$, and
$\psi(2S)\ar\gamma\chi_{c1},\gamma\chi_{c2}\ar\gamma\gamma\jpsi$.
  The decay $J/\psi \to \omega \pi ^+\pi ^-$ is studied.
  At low $\pi \pi$ mass, a large, broad peak due to the $\sigma$ is
  observed, and its pole position is determined.
Results are presented on $\psi(2S)$ and $J/\psi$ hadronic decays to
$K^0_SpK^-\bar n$ and $K^0_S\bar p K^+n$ final states.
No significant $\Theta(1540)$ signal, the
pentaquark candidate, is observed, and upper limits are set.
An enhancement near the $m_p + M_{\Lambda}$ mass threshold is
observed in the $p\bar{\Lambda}$ invariant mass
spectrum from $J/\psi \rightarrow p K^- \bar{\Lambda} + c.c.$ decays.
It can be fit with an S-wave Breit-Wigner resonance with a mass
$m=2075\pm 12 \:({\rm stat}) \pm 5 \:({\rm syst})$~MeV and a width of
$\Gamma =90 \pm 35 \:({\rm stat}) \pm 9 \:({\rm syst})$~MeV.

\keywords{charmonium; pentaquark; hadronic transitions.}
\end{abstract}

\section{Introduction}	

The Beijing Spectrometer (BES) is a general purpose solenoidal
detector at the Beijing Electron Positron
Collider (BEPC).
BEPC operates in the center of mass energy range from 2 to 5 GeV
with a luminosity at the $J/\psi$ energy of approximately
$ 5 \times 10^{30}$ cm$^{-2}$s$^{-1}$.  BES (BESI) is described in
detail in Ref. \refcite{bes1}, and the upgraded BES detector (BESII) is
described in Ref. \refcite{bes2}.

\section{$B(J/\psi \rt \pi^+ \pi^- \pi^0$)}

The largest $J/\psi$ decay involving hadronic resonances is $J/\psi
 \rt \rho(770) \pi$.  Its branching fraction has been reported by many
 experimental
 groups~\cite{PDG04}
 assuming all $\pi^+ \pi^- \pi^0$ final states come from $ \rho(770)
 \pi$.
The precision of these measurements varies from $13\%$ to 25\%.
Here, we present two independent measurements of this branching
fraction using $\jpsi$ and $\psip$
decays.

\subsection{\boldmath  Absolute measurement of $\jpsi\rar\ppp$ decays}
Events with two oppositely charged tracks and at least two good
photons are selected. A 5-constraint (5C) kinematic fit is made
under the $\pi^+\pi^-\gamma\gamma$ hypothesis with the invariant mass of the
two photons being constrained to the $\pi^0$ mass, 
and the fit $\chi^{2}_{\ppp}$ is required to be less than 15.
After these requirements, 219691 $\ppp$ candidates are selected.
The branching fraction is
$B(\jpsi\ar\ppp)=(21.84\pm0.05\pm 2.01)\times 10^{-3}$,
where the first error is statistical and the second systematic.

\begin{figure}[htbp]
\centerline{\epsfig{figure=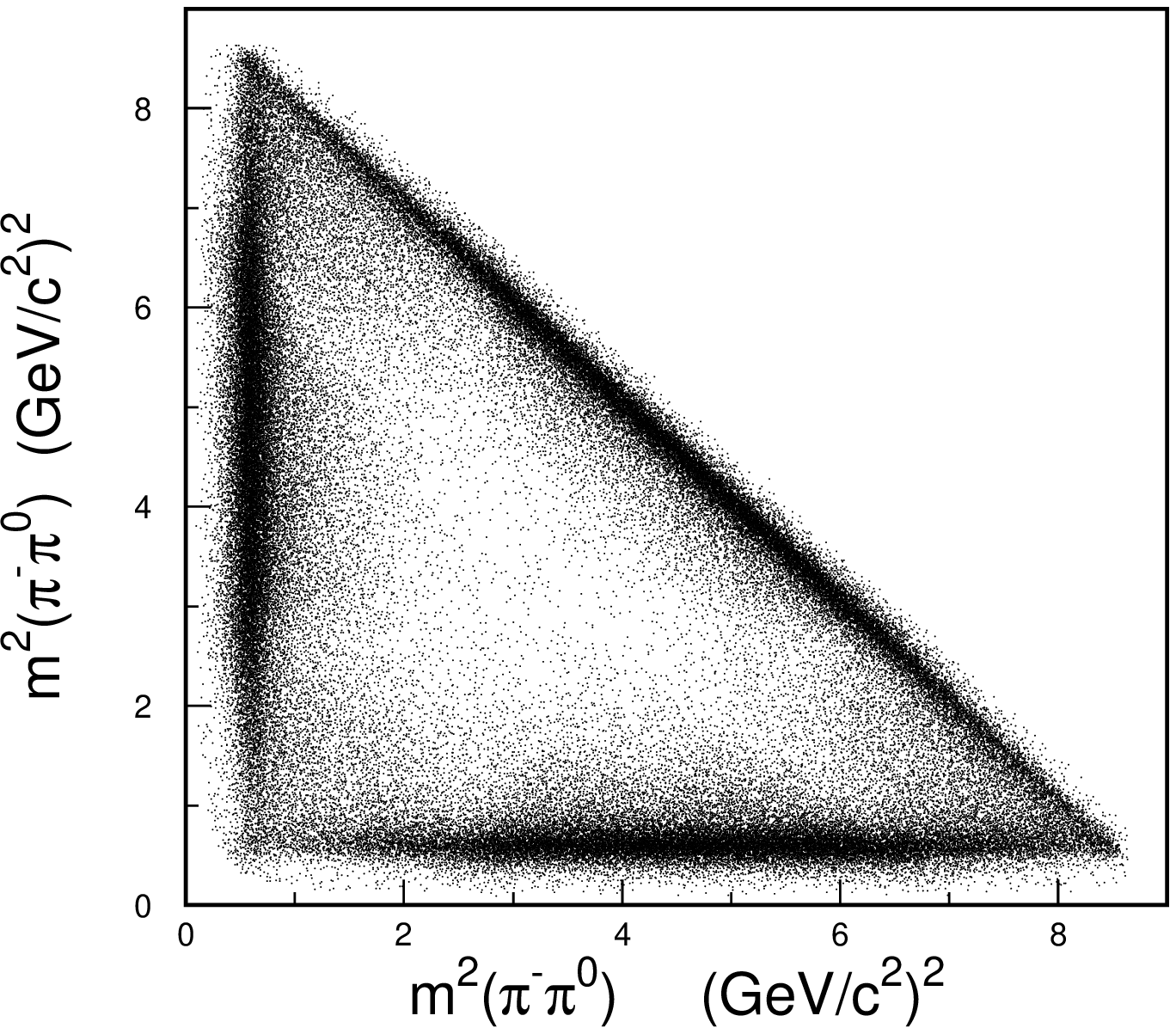,height=4.5cm}}
\caption{\label{dalitz}
  The Dalitz plot for $\jpsi\to\ppp$.}
\end{figure}

The Dalitz plot of $m_{\pi^+\pi^0}$ versus $m_{\pi^-\pi^0}$ is shown in
Fig. \ref{dalitz}. Three  bands are clearly visible in the plot, 
 corresponding to $\jpsi\to\rho\pi$;
$\jpsi\ar\ppp$ is
strongly dominated by $\rho\pi$.
 
\subsection{\boldmath Relative measurement of $\jpsi\ar\ppp$}
 The relative measurement is based on a sample of 14 million $\psi(2S)$ events. 
The $\psi(2S)$ is a copious source of
$J/\psi$ decays: the branching fraction of $\psi(2S)\rightarrow\pi^+\pi^-
J/\psi$ is the largest single $\psi(2S)$ decay channel. Therefore,
we can determine the branching fraction of $J/\psi\rightarrow\pi^+\pi^-\pi^0$
from a comparison of the following two processes:
\begin{eqnarray*}
\psi(2S)\rightarrow\pi^+\pi^- &J/\psi& \\
                     &\hookrightarrow &\pi^+\pi^-\pi^0 ~~~~~~~~~~~~~~~~(I) \\
 {\rm and}                 &\hookrightarrow & \mu^+\mu^- ~~~~~~~~~~~~~~~~~(II)
\end{eqnarray*}
Using the relative measurement, many
systematic errors mostly cancel. Therefore, the
precision of the branching fraction $\jpsi\rar\pi^+\pi^-\pi^0$ from
the relative measurement is
comparable with that of the direct $\jpsi$ decay, although the size of $\psip$ sample is smaller.

For process I, 
a 5C kinematic fit is performed for each
candidate event, and
 the event probability given by the fit must be greater than 0.01.
A 4C kinematic fit is performed for
$\psi(2S)\rightarrow\pi^+\pi^-\mu^+\mu^-$ candidate events, and
 the probability given by the fit must be greater than 0.01.
The branching fraction is
$B(J/\psi\rightarrow\pi^+\pi^-\pi^0)= (20.91\pm0.21\pm1.16)\times 10^{-3}.$


The results of the two measurements are in good agreement.
Their weighted mean is
$$B(J/\psi\rightarrow\pi^+\pi^-\pi^0)=(2.10\pm0.12)\% .$$
The result obtained is higher than those of previous
measurements and has better precision. For more detail, see Ref. \refcite{3pi}.

\section{$J/\psi \rt \omega \pi^+ \pi^-$}

There has been evidence for a low mass pole in the early DM2
\cite{dm2} and BESI \cite{bessig} data on $J/\psi \to \omega \pi ^+ \pi
^-$.  Here results on $J/\psi \to \omega \pi ^+ \pi ^-$ from $5.8 \times 10^7
J/\psi $ events collected with the BESII detector are
presented.

\begin{figure}[htbp]
\begin{center}
\includegraphics[width=10.0cm]{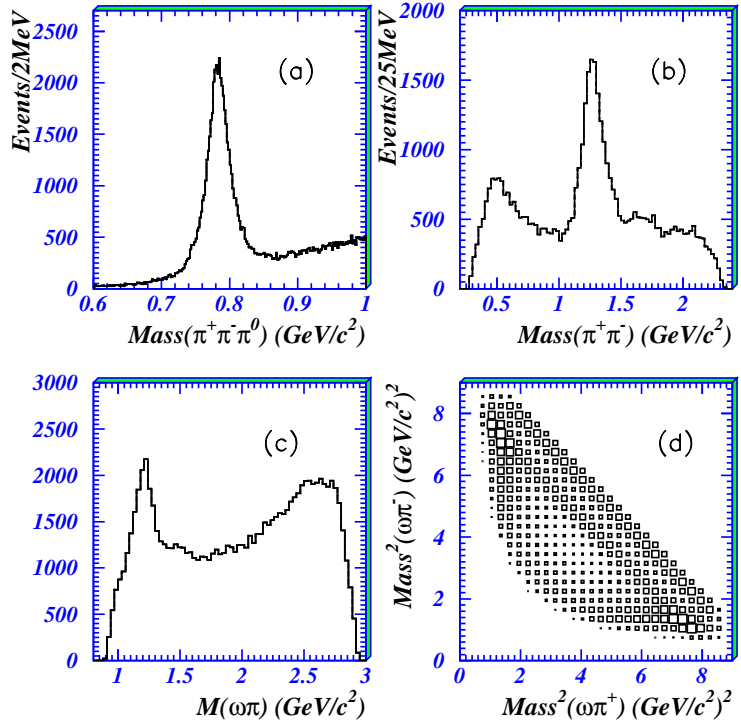}
\caption[]{\label{fig:sigma} $J/\psi \to \omega \pi ^+ \pi ^-$ (a.) Distribution of $\pi ^+\pi ^- \pi ^0$ mass.
  (b.) Distribution of the $\pi^+ \pi^-$ invariant mass recoiling
  against the $\omega$. (c.) Distribution of $\omega \pi$ invariant mass.
  (d.) Dalitz plot.}
\end{center}
\end{figure}

The $\omega$ is observed in its $\pi ^+ \pi ^- \pi ^0$ decay
mode.  Events are required to have four good charged
tracks with total charge zero and more than one good photon.  The TOF
and $dE/dx$ information are used to identify pions; they largely
reject kaons from background reactions such as $K^+K^-\pi
^+\pi^-\pi^0$.
Events with a $2\gamma$ invariant mass $|M_{\gamma \gamma }
- M_{\pi ^0}| < 40$ MeV/$c^2$ are fitted with a 5C kinematic fit to
$\pi ^+\pi ^-\pi ^+\pi ^- \pi ^0$ with the two photons being
constrained to the $\pi^0$ mass. Events with $\chi^2_{5C} <40$ are
selected.  The resulting $\pi ^+\pi ^- \pi ^0$ mass distribution for
is shown in Fig.~\ref{fig:sigma}(a). The $\omega$ signal is selected by requiring
$|M_{\pi ^+ \pi ^- \pi ^0} - M_{\omega }| \le 40$ MeV/$c^2$.

Fig.~\ref{fig:sigma}(b) shows the $\pi^+ \pi^-$ invariant mass spectrum which
recoils against the $\omega$, and Fig.~\ref{fig:sigma}(c) shows the $\omega \pi$
invariant mass. The Dalitz plot of this channel is shown in Fig.~\ref{fig:sigma}(d).
There is a large $f_2(1270)$ peak in Fig.~\ref{fig:sigma}(b) and a strong
$b_1(1235)$ peak in Fig.~\ref{fig:sigma}(c).  At low $\pi \pi $ masses in Fig.~\ref{fig:sigma}(b),
a broad enhancement which is due to the $\sigma$ pole is clearly seen.
This peak is evident as a strong band along the upper right-hand edge
of the Dalitz plot in Fig.~\ref{fig:sigma}(d). 

Partial wave analyses (PWA) are performed on this channel using two
methods.  In the first method, the whole mass region of $M_{\pi^+
  \pi^-}$ which recoils against the $\omega$ is analyzed, the $\omega$
decay information is used, and the background is subtracted by
sideband estimation. For the second method, the region $M_{\pi^+
  \pi^-} < 1.5$ GeV is analyzed, and the background is fitted by
$5\pi$ phase space.  In both methods, different parameterizations of
the $\sigma$ pole are also studied.

The mass and width of
the $\sigma$ are
different when using different $\sigma$ parameterizations. However, the pole
position of the $\sigma$ is stable; different analysis methods
and different parameterizations of the $\sigma$ amplitude give consistent
results for the $\sigma$ pole.
From a simple mean of the six analyses,
the pole position of the $\sigma$ is determined to be
$(541 \pm 39$ - $i$ $(252 \pm 42))$ MeV. Here, the errors cover
the statistical and systematic errors in the six analyses, as well as the
error in the extrapolation to the pole. The systematic errors dominate.
More detail may be found in Ref.~\refcite{sigma}.

\section{$\psi(2S) \rt \gamma \gamma J/\psi$ and $\psi(2S) \rt X J/\psi$}

Experimental results for the processes $\psip\ra \pi^0\J$, $\eta \J$, and
$\gamma\chi_{c1,2}$ are few and were mainly taken in the 1970s and
80s.~\cite{PDG04} 
Here, we report on the analysis of $\psip\ra\pi^0\J$,
$\eta\J$, and $\gamma\chi_{c1,2}$ decays based on a sample of
$14.0\times 10^6$ $\psip$ events collected with the BESII detector.
Events with two charged tracks identified as an electron pair or muon
pair and two or three photon candidates are selected.  A five
constraint (5C) kinematic fit to the hypothesis $\psip\ra\ggll$ with
the invariant mass of the lepton pair constrained to $\J$ mass is
performed, and the fit probability is required to be greater than
0.01.
 
\begin{figure}[htbp]
\hspace{0.2cm}
\centerline{\includegraphics[height=4.cm,width=7.5cm]{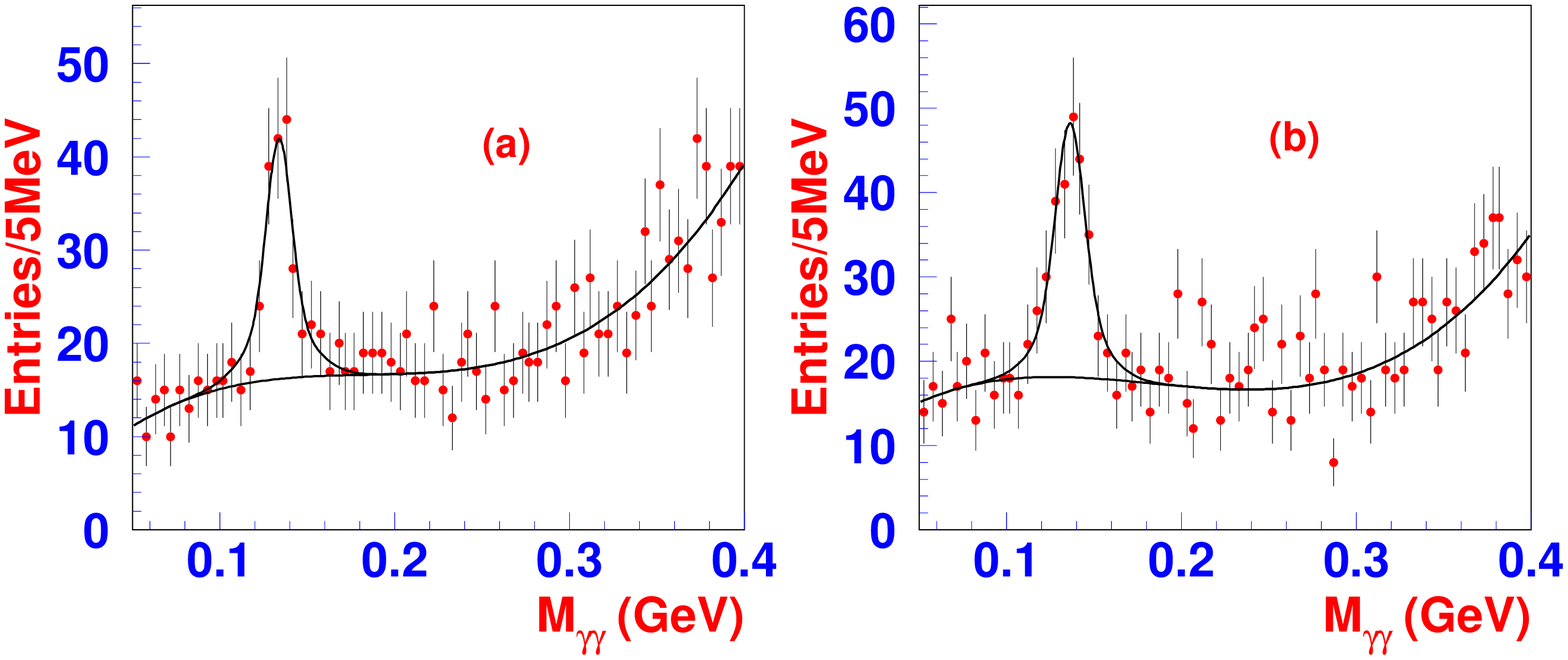}}
\caption{\label{fig:MJpiee}
  Two photon invariant mass distribution for candidate $\psip\ra\Jpi$ 
events for (a) $\ggee$ and (b) $\gguu$.}
\end{figure}

Fig.~\ref{fig:MJpiee} shows, after a cut
to remove the huge background from $\psip\ra\gx$ under the
$\psip\ra\pi^0\J$ signal,
the distribution of invariant mass, $\Mgg$. A Breit Wigner
with a double Gaussian mass resolution function to describe the
$\pi^0$ resonance plus a background polynomial is fitted
to the data. Similar analyses are made for the other channels.

\begin{table}[htbp]
\doublerulesep 0.5pt
\tbl{\label{results} Results from $\psi(2S) \rt \gamma \gamma J/\psi$}
{
\begin{tabular}{c|cc|cc}              \hline        \hline
Channel&\multicolumn{2}{c}{$\pi^0\jpsi$}&\multicolumn{2}{c}{$\eta\jpsi$}\\\hline
Final state&$\ggee$&$\gguu$&$\ggee$&$\gguu$\\\hline
B (\%)&$0.139\pm 0.020\pm 0.013$&$0.147\pm 0.019\pm
0.013$&$ 2.91\pm 0.12\pm 0.21$&$3.06\pm 0.14\pm 0.25 $\\
Combined (\%)&\multicolumn{2}{|c|}{$0.143\pm 0.014\pm 0.013$}
&\multicolumn{2}{|c}{$2.98\pm 0.09\pm 0.23$}\\
PDG (\%)\cite{PDG04}&\multicolumn{2}{|c|}{$0.096\pm 0.021 $}&\multicolumn{2}{|c}
{$3.16\pm 0.22$}\\\hline\hline
Channel&\multicolumn{2}{c}{$\gamma\chi_{c1}$}&\multicolumn{2}{c}
{$\gamma\chi_{c2}$}\\\hline
Final state&$\ggee$&$\gguu$&$\ggee$&$\gguu$\\\hline
B (\%)&$8.73\pm 0.21\pm 1.00$&$9.11\pm 0.24\pm 1.12$&
$7.90\pm 0.26\pm 0.88$&$8.12\pm 0.23\pm 0.99$\\
Combined (\%)&\multicolumn{2}{|c|}{$8.90\pm 0.16\pm
1.05$}&\multicolumn{2}{|c}{$8.02\pm 0.17\pm 0.94$}\\
PDG (\%)\cite{PDG04}&\multicolumn{2}{|c|}{$8.4\pm 0.6$}&\multicolumn{2}{|c}
{$6.4\pm 0.6$}\\\hline
\end{tabular}
}
\end{table}

Using the fitting results and the efficiencies and correction factors
for each channel, the branching fractions listed in
Table~\ref{results} are determined. The BES $B(\psip\ra\pi^0\J)$
measurement has improved precision by more than a factor of two
compared with other experiments, and the $\psip\ra\eta\J$ branching
fraction is the most accurate single measurement. More details on this
analysis may be found in Ref.~\refcite{ggJ}.


In another analysis, based on a sample of approximately $4 \times
10^6$ $\psip$ events obtained with the BESI detector,~\cite{besI} a
different technique is used for measuring branching fractions for the
inclusive decay $\psip \rt {\rm anything} J/\psi $, and the exclusive
processes for the cases where $X = \eta$ and $X = \pi\pi$, as well as
the cascade processes $\psip \rt \gamma \chi_{c0/1/2}$, $\chi_{c0/1/2}
\rt \gamma J/\psi$.  Inclusive $\mu^+ \mu^-$ pairs are reconstructed,
and the number of $\psip \rt J/\psi X$ events is determined from the
$J/\psi \rt \mu^+ \mu^-$ peak in the $\mu^+ \mu^-$ invariant mass
distribution.  The exclusive branching fractions are determined from
fits to the distribution of masses recoiling from the $J/\psi$ with
Monte-Carlo determined distributions for each individual channel.



\begin{figure}[!htb]
\begin{center}
\begin{minipage}[t]{2.4in}
\centerline{\epsfysize 6 cm
\epsfbox{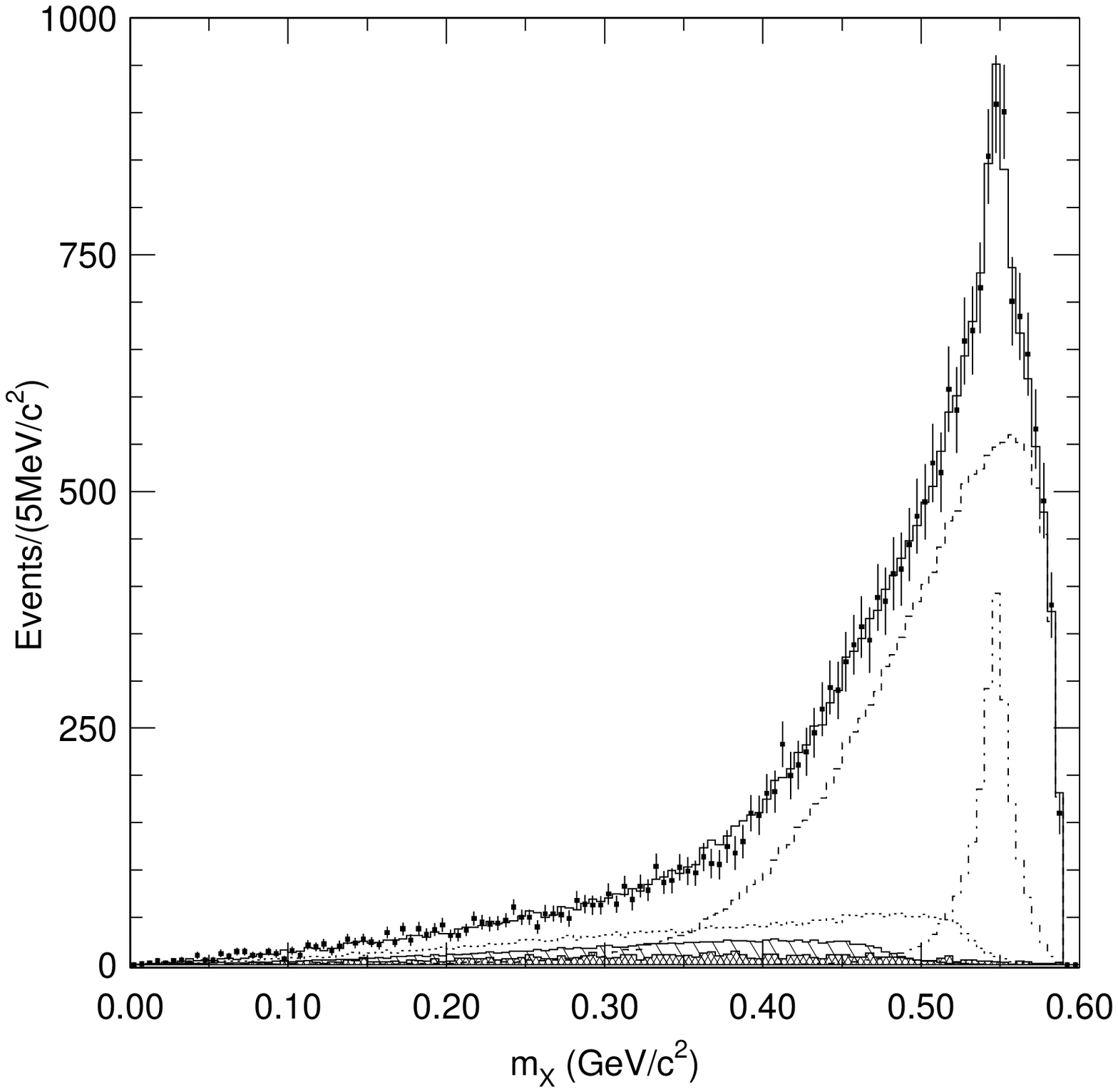}}
\caption{\label{fig:none} Fit of the $m_X$ distribution events with
no additional charged tracks.  Shown are the data (points with error 
bars), the component histograms, and the final fit.  For the components, 
the large, long-dash histogram is $\psi(2S) \rt J/\psi \pi \pi$, the
narrow, dash-dot histogram is $\psi(2S) \rt J/\psi \eta $, the broad,
short-dashed histogram is $\psip \rt \gamma \chi_{c1},  \chi_{c1} \rt
\gamma J/\psi$, the broad,
hatched histogram is $\psip \rt \gamma \chi_{c2},  \chi_{c2} \rt \gamma J/\psi$, and the lowest
cross-hatched histogram is the combined $e^+ e^- \rt \gamma \mu^+
\mu^-$ and $e^+ e^- \rt \psi(2S), \psi(2S) \rt (\gamma)\mu^+ \mu^-$
background. The final fit is the solid histogram.
}
\end{minipage} \ \
\begin{minipage}[t]{2.40in}
\centerline{\epsfysize 6 cm
\epsfbox{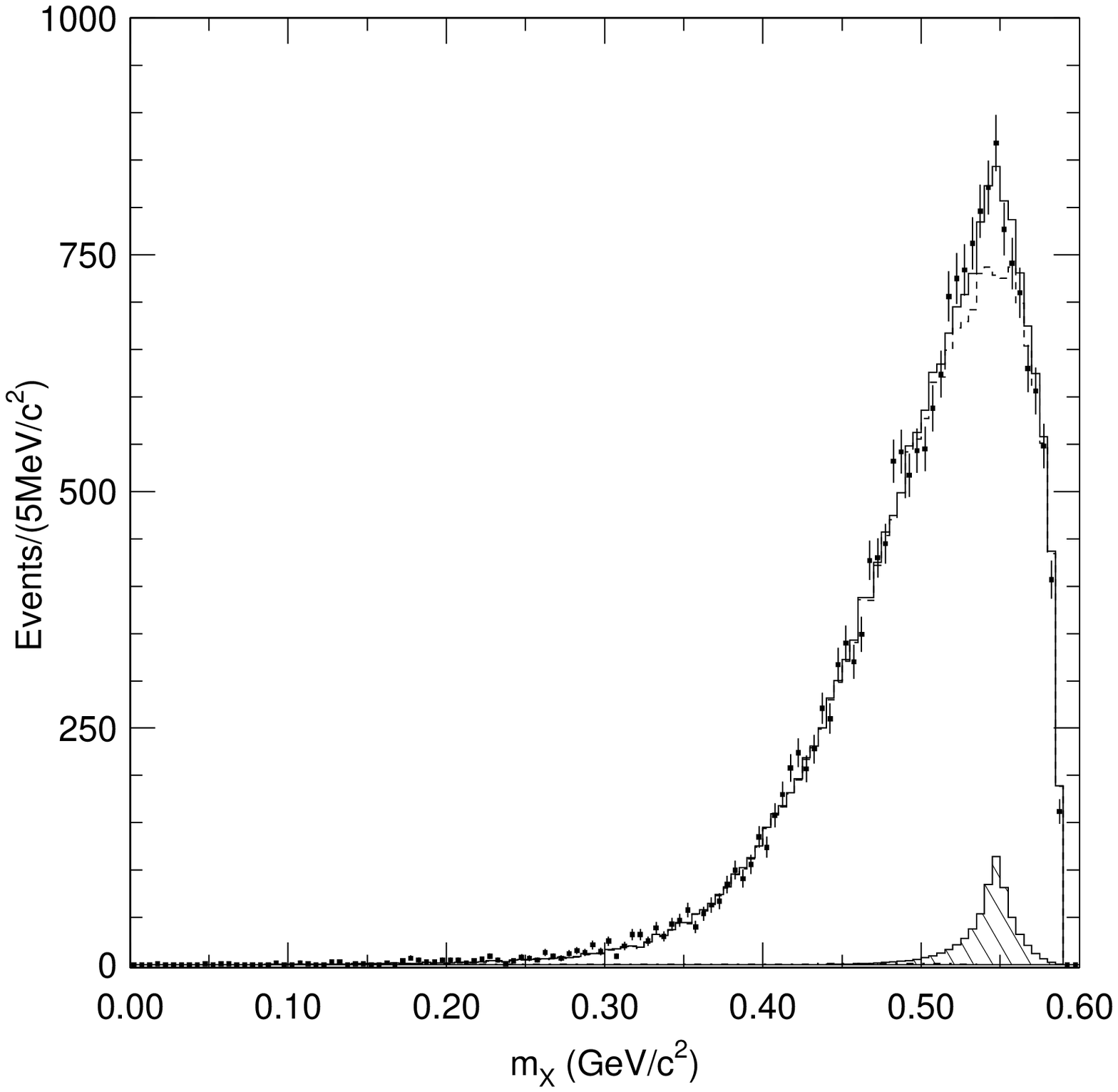}}
\caption{\label{fig:one} Fit of the $m_X$ distribution for events with
any number of additional charged tracks.  Shown are the data (points
with error bars), the component histograms, and the final fit (solid
histogram). The dashed histogram is $\psi(2S) \rt J/\psi \pi^+ \pi^-$,
and the hatched histogram is $\psi(2S) \rt J/\psi \eta$.  There is
very little evidence for $\psi(2S) \rt \gamma \chi_{c1/2},
\chi_{c1/2} \rt \gamma J/\psi$.  This distribution is composed
predominantly of $\psi(2S) \rt J/\psi \pi^+ \pi^-$.  }
\end{minipage}
\end{center}
\end{figure}

The mass recoiling against the $J/\psi$ candidates, $m_X$ is
determined from energy and momentum conservation. To determine the
number of exclusive decays and separate $\psi(2S) \rt J/\psi \pi^0
\pi^0$ and $\psi(2S) \rt J/\psi \pi^+ \pi^- $ events, $m_X$ histograms
for events with and without additional charged tracks, shown in
Figs.~\ref{fig:none} and \ref{fig:one}, are fit simultaneously.
To reduce background and improve the quality of the track momentum
measurements, events used for this part of the analysis are required
to have a kinematic fit $\chi^2 < 7$.  
The channels of interest are
normalized to the observed number of $ \pi^+ \pi^-J/\psi$ events;
ratios of the studied branching fractions to that for $B(\psip
\rt  \pi^+ \pi^- J/\psi)$ are reported.  This has that advantage
that many of the systematic errors largely cancel.

\begin{table}[!h]
\tbl{\label{tab:results} Final branching ratios and branching
  fractions.  PDG04-exp results are single measurements or
  averages of measurements, while PDG04-fit are results of their
  global fit to many experimental measurements. The BES results in the second
  half of the table are calculated using the PDG04 value of $B_{\pi \pi}
  = B(\psip \rt J/\psi \pi^+ \pi^- ) = (31.7 \pm 1.1) \%$. }
{\begin{tabular} {|l|c|c|c|} \hline
Case                                      &   This result &
PDG04-exp & PDG04-fit  \\ \hline
$B(J/\psi \: ({\rm anything})/B_{\pi \pi}$   & $1.867  \pm 0.026  \pm 0.055$ & $2.016
\pm 0.150$  &$1.821 \pm 0.036$   \\
$B(J/\psi\pi^0 \pi^0)/B_{\pi \pi}$ & $0.570 \pm 0.009 \pm
0.026$ & - & $0.59 \pm 0.05$ \\
$B(J/\psi \eta)/B_{\pi \pi}$ & $0.098 \pm  0.005 \pm
0.010$ & $0.091 \pm 0.021$ & $0.100 \pm 0.008$\\ 
$B(\gamma \chi_{c1})B(\chi_{c1} \rt \gamma J/\psi)/B_{\pi \pi}$ & $0.126
\pm 0.003 \pm 0.038$ & $0.085 \pm 0.021$  & $0.084 \pm 0.006$ \\ 
$B(\gamma \chi_{c2})B(\chi_{c2} \rt \gamma J/\psi)/B_{\pi \pi}$ & $0.060
\pm 0.000 \pm 0.028$ & $0.039 \pm 0.012$  & $0.041 \pm 0.003$ \\ \hline
$B(J/\psi \: ({\rm anything})$ (\%)  & $59.2  \pm 0.8 \pm 2.7  $ & $55 \pm 7$ & $57.6 \pm 2.0$ \\ 
 $B(J/\psi \pi^0 \pi^0)$ (\%)  &  $18.1 \pm 0.3 \pm 1.0 $ & -- & $18.8 \pm 1.2$\\
 $B(J/\psi \eta)$  (\%)    &  $3.11 \pm 0.17 \pm 0.31 $ & $2.9 \pm
0.5$ & $3.16 \pm 0.22$\\ 
 $B(\gamma \chi_{c1})B(\chi_{c1} \rt \gamma J/\psi)$ (\%)  & $4.0 \pm
 0.1 \pm 1.2$ & $2.66 \pm 0.16$& $2.67 \pm 0.15$\\ 
 $B(\gamma \chi_{c2})B(\chi_{c2} \rt \gamma J/\psi)$ (\%)  & $1.91 \pm 0.01 \pm 0.86$
& $1.20 \pm 0.13$& $1.30 \pm 0.08$\\ \hline
\end{tabular}}
\end{table}

The final branching fraction ratios and branching fractions are shown
in Table~\ref{tab:results}, along with the PDG
experimental averages and global fit results.  The results for $B(J/\psi
\:{\rm anything})/B(\psi(2S) \rt  \pi^+ \pi^- J/\psi)$ and $B(\eta J/\psi)/B(\psi(2S) \rt  \pi^+ \pi^- J/\psi)$ have smaller errors than
the previous results.
The agreement for
both the ratios of branching fractions and the calculated branching
fractions using the PDG result for $\psip \rt B( \pi^+ \pi^-J/\psi)$
with the PDG fit results is good, and the determination of $B(
 \eta J/\psi)$ agrees well with the determination from $\psi(2S)  \rt \gamma
\gamma J/\psi$ decays above.
 More details on this
analysis may be found in Ref.~\refcite{Xjpsi}.

\section{Pentaquark Search}

The pentaquark,\cite{LEPS} the main topic of the opening session of
this conference, has generated much excitement.  BES has searched for
the pentaquark state $\Theta(1540)$ in $\psi(2S)$ and $J/\psi$ decays
to $K^0_SpK^-\bar n$ and $K^0_S\bar p K^+n$ final states using samples
of 14 million $\psi(2S)$ and 58 million $J/\psi$ events taken with
BES\,II. These processes could contain
$\Theta$ decays to $K^0_Sp,~K^+n~(uudd\bar s)$ and $\bar\Theta$ decays
to $K^0_S\bar p,~K^-n~(\bar u\bar u\bar d\bar d s)$.

The anti-neutron and neutron are not detected. The $K^0_S$ meson in
the event is identified through the decay $K^0_S\to\pi^+\pi^-$.
Candidate events are
kinematically fitted under the assumption of a missing $\bar n(n)$ to
obtain better mass resolution and to suppress the backgrounds. 
Events with missing mass close to the $\bar n (n)$'s
mass are selected.  We use the same criteria and treatment for both
$\psi(2S)$ and $J/\psi$ data.

\begin{figure}[htbp]
\begin{center}
\epsfxsize=4.5cm\epsffile{plot4.epsi}
\vspace*{5pt}
\caption{\label{fig:penta}Scatter plot of $K^-n~(K^0_S \bar p)$  versus $K^0_S p~(K^+ n)$
for
$\psi(2S)\to K^0_SpK^-\bar n$ +
$K^0_S\bar p
K^+n$ modes.}
\end{center}
\end{figure}

The scatter plot of $K^-n~(K^0_S \bar p)$ versus $K^0_S p~(K^+ n)$ for
$\psi(2S)\to K^0_SpK^-\bar n$ + $K^0_S\bar p K^+n$ modes is shown in
Fig.~\ref{fig:penta}.
Zero events fall within the signal region, shown as a square centered
at $(1.540, 1.540)$ GeV/c$^2$, and we set an upper limit
 at the 90\% confidence level (C.L.) on the branching ratio:
$${\cal B}(\psi(2S)\to\Theta\bar\Theta\to
K^0_S p K^-\bar n + K^0_S \bar p K^+ n)  <
0.84\times 10^{-5}.$$

Another possibility is that the $\psi(2S)$ decays to only one $\Theta$ or
$\bar\Theta$ state.  To determine the number of $\Theta(1540)$
events from single $\Theta$ or $\bar\Theta$ production, we count the
number of events within regions of 1.52 - 1.56 GeV$/c^2$ in the
projections of Fig.~\ref{fig:penta} and set upper limits, shown in Table~\ref{tab:penta}.

\begin{table}[htbp]
\tbl{\label{tab:penta} Summary of upper limits.}
{\begin{tabular}{l|c|c}
\hline
\hline
Decay mode~~&~~$\psi(2S)$~~~~& $J/\psi$ \\
\hline
$\to\Theta\bar\Theta\to K^0_S p K^-\bar nx+ K^0_S \bar p K^+
n$ & $0.88\times 10^{-5}$ & $1.1\times 10^{-5}$\\
$\to\Theta K^-\bar n\to K^0_S p
K^-\bar n$ & $1.0\times 10^{-5}$ & $2.1\times10^{-5}$\\
$\to\bar\Theta K^+ n\to K^0_S\bar p
K^+ n$ & $2.6\times 10^{-5}$ & $5.6\times10^{-5}$\\
$\to K^0_S p\bar\Theta \to K^0_S p
K^-\bar n$ & $0.60\times 10^{-5}$ &  $1.1\times10^{-5}$\\
$\to K^0_S\bar p\Theta\to K^0_S\bar p
K^+ n$ & $0.70\times 10^{-5}$ & $1.6\times10^{-5}$\\
\hline\hline
\end{tabular}
}
\end{table}

For the decays of $J/\psi\to K^0_SpK^-\bar n$ and
$K^0_S\bar p
K^+n$, we use the same
criteria and analysis method as those used for the $\psi(2S)$ data to study
possible $\Theta(1540)$ production.
There is no significant $\Theta(1540)$ signal, and we determine
upper
limits
on the
branching fractions at the 90\% C.L., shown in Table~\ref{tab:penta}. 
Full details may be found in Ref. \refcite{penta}.

\section{Enhancement in the $p \bar{\Lambda}$ invariant 
mass spectrum 
in $J/\psi \rightarrow p K^- \bar{\Lambda}$ and 
in $ \psi(2S) \rightarrow p K^- \bar{\Lambda}$ decays}

An anomalous enhancement near the mass threshold in the 
$p\bar{p}$ invariant mass
spectrum was observed by the BES II experiment in 
$\jpsi\rightarrow\gamma p\bar{p}$ decays.~\cite{gpp} 
This enhancement 
can be fitted with an S-wave Breit-Wigner resonance function
with a mass around 1860~MeV and a width $\Gamma<30$~MeV,
and has been interpreted as a possible baryonium state.~\cite{baryonium}
Similar  $p\bar{p}$ mass-threshold enhancements
have been observed
in the decays $B^+ \rightarrow K^+ \ppbar$ and
$\bar{B}^0 \rightarrow D^0 \ppbar$  by the 
Belle Collaboration.~ \cite{belle_pp1,belle_pp2}
It is, therefore, of special interest to search for possible resonant 
structures  in other baryon-antibaryon final states. 
The Belle Collaboration recently observed a near-threshold enhancement 
in  the $p\bar{\Lambda}$ mass spectrum from
$B \rightarrow p\bar{\Lambda}\pi$ decays.~\cite{belle_pL}
We report the observation of	
an enhancement near threshold in the 
$p \bar{\Lambda}$ invariant 
mass spectrum 
in $J/\psi \rightarrow p K^- \bar{\Lambda}$ and 
in $ \psi(2S) \rightarrow p K^- \bar{\Lambda}$ decays.
The results are based on an analysis of 
$5.8 \times 10^7$ $J/\psi$ and $1.4 \times 10^7$ $\psi(2S)$ decays 
detected  in BESII.

The $J/\psi \rightarrow pK^- {\bar \Lambda}$ candidate events 
are required to have four good charged tracks
with total charge zero.
Events
are subjected to a four-constraint (4C) kinematic fit with the
corresponding mass assignments for each track.
For events with ambiguous particle identification,  all possible  
4C combinations are formed, 
and the combination with the smallest $\chi^2$ is chosen.
A sample of  $5421$ 
$J/\psi \rightarrow pK^- {\bar \Lambda}$ candidates
survive the final selection. 
Monte Carlo studies indicate
that the background in the selected event sample is at the $1 \sim
2\%$ level.


	\begin{figure}[hbtp]
	\centerline{
	\epsfysize 4.2 truein
	\epsfbox{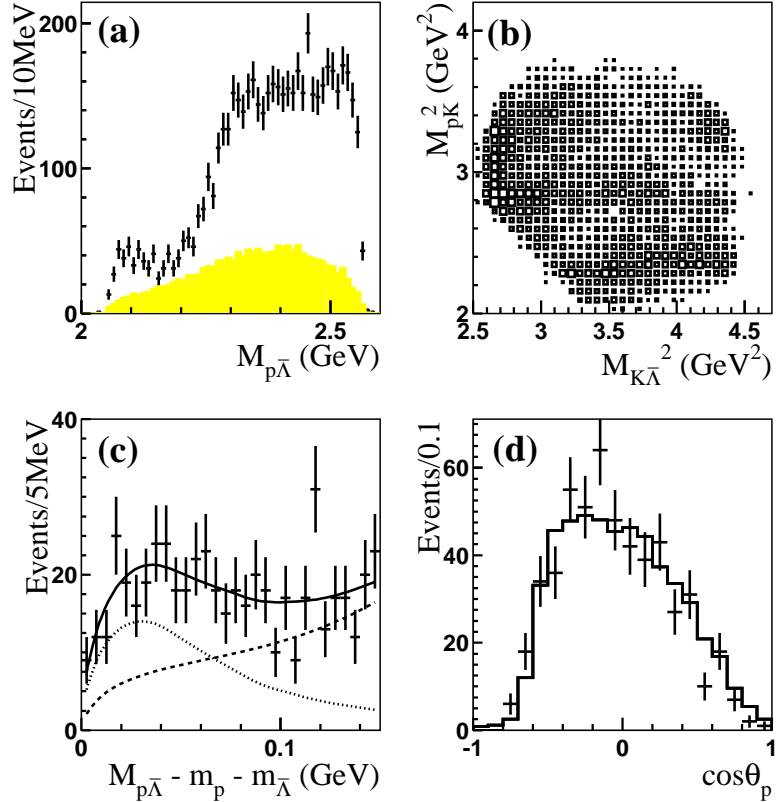}
	}
	\caption{ 
	    (a) The points with error bars indicate the
            measured $p\bar\Lambda$ mass spectrum;
	    the shaded histogram indicates  
            phase space MC events (arbitrary normalization).
	    (b) The Dalitz plot for the selected event sample.
	    (c) A fit (solid line) to the data. The 
	    dotted curve indicates the Breit-Wigner signal and the 
            dashed curve the 
	    phase space `background.'
	    (d) The $\cos \theta_p$ distribution under the enhancement,
                the points are data and the histogram is the MC
              (normalized to data)}
	\label{x208}
	\end{figure}

The $p\bar{\Lambda}$ invariant mass spectrum for the selected events
is shown in Fig.~\ref{x208}(a), where an enhancement is evident near
the mass threshold.  No corresponding structure is seen in a sample of
$J/\psi \rightarrow p K^- \bar{\Lambda}$ MC events generated with a
uniform phase space distribution.  The $pK^- \bar\Lambda$ Dalitz plot
is shown in Fig.~\ref{x208}(b).  In addition to bands for the well
established $\Lambda^*(1520)$ and $\Lambda^*(1690)$, there is a
significant $N^*$ band near the $K^- \bar{\Lambda}$ mass threshold,
and a $p\bar{\Lambda}$ mass enhancement, isolated from the $\Lambda^*$
and $N^*$ bands, in the right-upper part of the Dalitz plot.

This enhancement can be fit 
with an acceptance weighted S-wave Breit-Wigner function, 
together with a function  describing the phase space 
contribution, as shown in Fig.~\ref{x208}(c). 
The fit gives a peak mass of $m=2075\pm 12 \pm 5$~MeV and a width
 $\Gamma=90 \pm 35 \pm 9$~MeV. 
The
significance is about $ 7\sigma$. 
A P-wave Breit-Wigner resonance functions can also fit the
enhancement.
The $\cos\theta_p$ distribution, shown in Fig.~\ref{x208}(d), where $\theta_p$ is the
decay angle of p in the $p\bar{\Lambda}$ CM frame, agrees well with that of a MC
sample of $J/\psi\rightarrow K X \rightarrow K p \bar{\Lambda}$.  Since 
the MC $\cos\theta_p$ distribution is generated as a uniform S-wave
distribution and the detected MC distribution agrees with data
in Fig.~\ref{x208}(d), the observed distribution for the enhancement
is consistent with S-wave decays to $p \bar{\Lambda}$.

Evidence of a similar enhancement is observed in 
$ \psi(2S) \rightarrow p K^- \bar{\Lambda}$ when 
the same analysis is performed on the $\psi(2S)$ data sample.
More detail can be found in Ref.~\refcite{pklambda}.


\section*{Acknowledgments}

I wish to acknowledge the efforts of my BES colleagues on all the
results presented here.  I also want to thank the organizers
for the opportunity to present these results at MESON2004.


\begin{thebibliography}{0}

\bibitem{bes1} J. Z. Bai {\it et al.}, (BES Collab.), {\it Nuc. Inst. Meth.} {\bf
A344}, 319 (1994).

\bibitem{bes2}  J. Z. Bai {\it et al.}, (BES Collab.), {\it Nuc. Inst. Meth.} {\bf
A458}, 627 (2001).

\bibitem{PDG04}
S. Eidelman {\it et al}., (Particle Data Group), {\it Phys. Lett.} {\bf B592}, 1 (2004).

\bibitem{3pi} J. Z. Bai {\it et al.}, (BES Collab.), accepted by {\it
  Phys. Rev.} {\bf D}, hep-ex/0402013.

\bibitem{dm2} J.E. Augustin {\it et al.}, {\it Nucl. Phys.} B320, 1 (1989).

\bibitem{bessig} Ning Wu (BES Collab.), Proceedings of the XXXVIth
Rencontres de Moriond, Les Arcs, France, March 17-24, (2001).

\bibitem{sigma} M. Ablikim {\it et al.}, (BES Collab.), accepted by {\it
  Phys. Lett.} {\bf B}, hep-ex/0406038.

\bibitem{ggJ}  M. Ablikim {\it et al.}, (BES Collab.), accepted by {\it
  Phys. Rev.} {\bf D}, hep-ex/0403023.

\bibitem{besI} J.Z. Bai {\it et al.}, (BES Collab.), {\it Nucl. Inst. Meth.}
{\bf A344}, 319 (1994).

\bibitem{Xjpsi} M. Ablikim {\it et al.}, BES Collab., accepted by {\it
  Phys. Rev.} {\bf D}, hep-ex/0404020.


\bibitem{LEPS} T. Nakano {\it et al.} (LEPS Collab.),
Phys. Rev. Lett. {\bf 91}, 012002 (2003).

\bibitem{penta}M. Ablikim {\it et al.}, (BES Collab.), accepted by {\it
  Phys. Rev.} {\bf D}, hep-ex/0402012.

\bibitem{gpp} J.Z. Bai {\it  et al.},  (BES Collab.),
    {\it Phys. Rev. Lett.} {\bf 91}, 022001 (2003).

\bibitem{baryonium}Alakabha Datta, Patrick J. O'Donnell,
{\it Phys. Lett.} {\bf B567}, 273 (2003).

\bibitem{belle_pp1} K. Abe {\it  et al.}, (Belle Collab.),
 {\it Phys. Rev. Lett.} {\bf 88}, 181803 (2002).

\bibitem{belle_pL} M.Z. Wang {\it  et al.},  (Belle Collab.),
 {\it Phys. Rev. Lett.} {\bf 90}, 201802 (2003).

\bibitem{belle_pp2} K. Abe {\it  et al.}, (Belle Collab.), 
 {\it Phys. Rev. Lett.} {\bf 89}, 151802 (2002).

\bibitem{pklambda}M. Ablikim {\it et al.}, (BES Collab.), accepted by {\it
  Phys. Rev.  Lett.}, hep-ex/0405050.

\end{thebibliography}
\end{document}